\documentclass[showpacs,amsmath,amssymb,twocolumn,superscriptaddress,floatfix,prl]{revtex4-2}
\usepackage[dvips]{graphicx}
\usepackage{enumerate}
\usepackage{epsfig}
\usepackage{xcolor}
\usepackage[T1]{fontenc}
\usepackage{fullpage}
\usepackage{amsthm,amsfonts,amscd,mathrsfs,xspace,framed}
\usepackage{color}
\usepackage{setspace}
\usepackage{url}
\usepackage{enumitem}

\begin{document}

\title{Control spontaneous symmetry breaking of photonic chirality with reconfigurable anomalous nonlinearity}

\author{Chaohan Cui}
\affiliation{James C. Wyant College of Optical Sciences, The University of Arizona, Tucson, Arizona 85721, USA}
\author{Liang Zhang}
\affiliation{James C. Wyant College of Optical Sciences, The University of Arizona, Tucson, Arizona 85721, USA}
\author{Linran Fan}
\email{lfan@optics.arizona.edu}
\affiliation{James C. Wyant College of Optical Sciences, The University of Arizona, Tucson, Arizona 85721, USA}

\maketitle

\textbf{
Spontaneous symmetry breaking in nonlinear systems provides a unified method to understand vastly different phenomena, ranging from Higgs mechanism~\cite{bernstein1974spontaneous} and superconductivity~\cite{baskaran1988gauge} to ecological stability~\cite{borile2012spontaneously} and genome generation~\cite{takeuchi2017origin}. Spontaneous symmetry breaking is typically considered as the intrinsic property of nonlinear systems with fixed occurrence condition and property, as the form and magnitude of nonlinear interactions cannot be modified~\cite{malomed2013spontaneous}. Here, we report the development of reconfigurable Kerr optical nonlinearity to control spontaneous symmetry breaking.
This is achieved through the interference between the intrinsic Kerr and cascaded second-order nonlinear processes~\cite{cui2022situ}. 
Anomalous Kerr effects including negative self-phase modulation and strength tuning between competing nonlinear processes have been demonstrated. With the reconfigurable Kerr nonlinearity, we realize the \textit{in-situ} prohibition and facilitation of spontaneous symmetry breaking of photonic chirality. This work could empower the experimental study of spontaneous symmetry breaking in unexplored regimes and inspire the development of novel photonic functions.}

The chiral degree of freedom can provide the unique propagation-direction-dependent property of photons, which is different from conventional photonic functions based on spatial~\cite{efron1994spatial,zhao2015capacity}, polarization~\cite{han2005coherent,crespi2011integrated}, and spectral-temporal~\cite{fan2016integrated,lukens2017frequency,fan2019spectrotemporal,raymer2020temporal} operations. Photonic chirality is typically introduced by explicit asymmetries including deformed photonic structures~\cite{redding2012local,cao2015dielectric}, gain-loss modulation in non-Hermitian systems~\cite{miao2016orbital,feng2017non,el2018non}, and interaction with magnetic materials~\cite{decker2007circular,liu2014spontaneous}. Such explicit symmetry breaking of photonic chirality has been used to build optical isolators~\cite{sayrin2015nanophotonic,xia2018cavity,sohn2021electrically}, realize topological photonic states~\cite{raghu2008analogs,hafezi2011robust,zhao2019non}, and develop unidirectional micro-lasers~\cite{peng2016chiral,carlon2019optically} in the classical regime, as well as improve single-photon sources~\cite{mitsch2014quantum, sollner2015deterministic, cui2021photonic} and develop logic operations in the quantum regime~\cite{junge2013strong,shomroni2014all,lodahl2017chiral}.

Besides explicit approaches, the symmetry breaking of photonic chirality can also happen spontaneously with Kerr nonlinearity~\cite{cao2017experimental,del2017symmetry,garbin2020asymmetric,cao2020reconfigurable,silver2021nonlinear,xu2021spontaneous}. This is introduced by the imbalance between self- and cross-phase modulation, two fundamental Kerr nonlinear effects. In this case, photons can show propagation-direction-dependent property even though the system configuration is completely symmetric. Unlike explicit approaches where the symmetry breaking can be engineered conveniently by photonic structures, the capability to control the condition of spontaneous symmetry breaking remains exclusive. The key challenge is the direct modification of Kerr nonlinearity, which typically has fixed form and magnitude determined by material intrinsic properties~\cite{gaeta2019photonic}.

Here, we report the approach to break the intrinsic restriction of Kerr nonlinearity, and demonstrate the controlled spontaneous symmetry breaking of photonic chirality. 
This is realized through the cascaded second-order nonlinearity, which is equivalent to reconfigurable artificial Kerr nonlinearity. In particular, we can selectively control the self-phase modulation coefficient without affecting the cross-phase modulation. The capability to tune the relative strength between self- and cross-phase modulation enables the modification of the spontaneous symmetry breaking condition for photonic chirality.

\begin{figure*}[htbp]
\centering
\includegraphics[width=\linewidth]{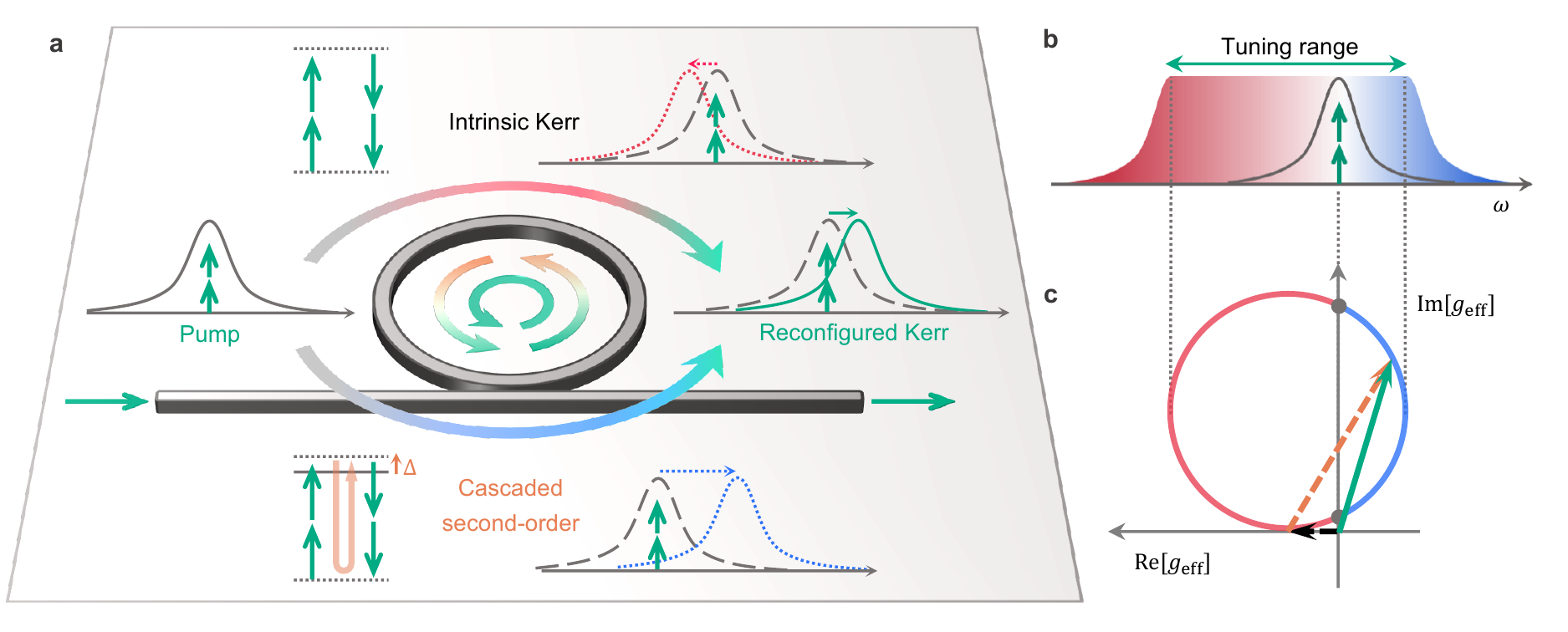}
\caption{\textbf{Schematics for reconfigurable Kerr nonlinearity.} \textbf{a}, Interference between the intrinsic Kerr and the cascaded second-order nonlinear processes in the photonic ring cavity. The intrinsic Kerr nonlinearity introduces resonance shift towards lower frequency (red-shift). This can be modeled as four-wave mixing with two pump photons (green) annihilated to re-generate two photons at the same frequency. The cascaded second-order nonlinearity introduces resonance shift towards higher frequency (blue-shift) with positive cavity detune. This can be modeled as artificial four-wave mixing, where two pump photons (green) are first combined to generate one second-harmonic photon (orange), which drives the parametric down-conversion to re-generate two photons at the pump frequency. The overall frequency shift is the coherent combination of the two parallel processes. 
\textbf{b}, Tuning range of the frequency shift induced by self-phase modulation with anomalous Kerr nonlinearity.
\textbf{c}, The vector representation of the effective Kerr nonlinear coefficient $g_{\rm eff}$. By tuning the cavity detune $\Delta$, the effective Kerr nonlinearity can change along the circle to give frequency shift towards lower (red) and higher (blue) frequency. Black dashed arrow: intrinsic Kerr nonlinearity; orange dashed arrow: cascaded second-order nonlinearity; green solid arrow: effective Kerr nonlinearity.}
\label{fig:Fig1} 
\end{figure*}

Our scheme for reconfigurable Kerr nonlinearity can be implemented in an integrated photonic ring cavity with second-order nonlinearity. We consider the cavity response to a continuous-wave pump field. The pump field is subject to two parallel nonlinear processes. The first one is the conventional self-phase modulation due to the intrinsic Kerr nonlinearity (top path in Fig.~\ref{fig:Fig1}a). The cavity instantaneous resonance is shifted by $\delta\omega\propto-g_3 \cdot N_\mathrm{p}$, with $g_3$ the Kerr nonlinear coefficient and $N_\mathrm{p}$ the intra-cavity pump photon number. As all photonic materials have positive Kerr nonlinear coefficients~\cite{gaeta2019photonic}, intrinsic Kerr nonlinearity causes the resonance shift towards the lower frequency. 

\begin{figure*}[htbp]
\centering
\includegraphics[width=\linewidth]{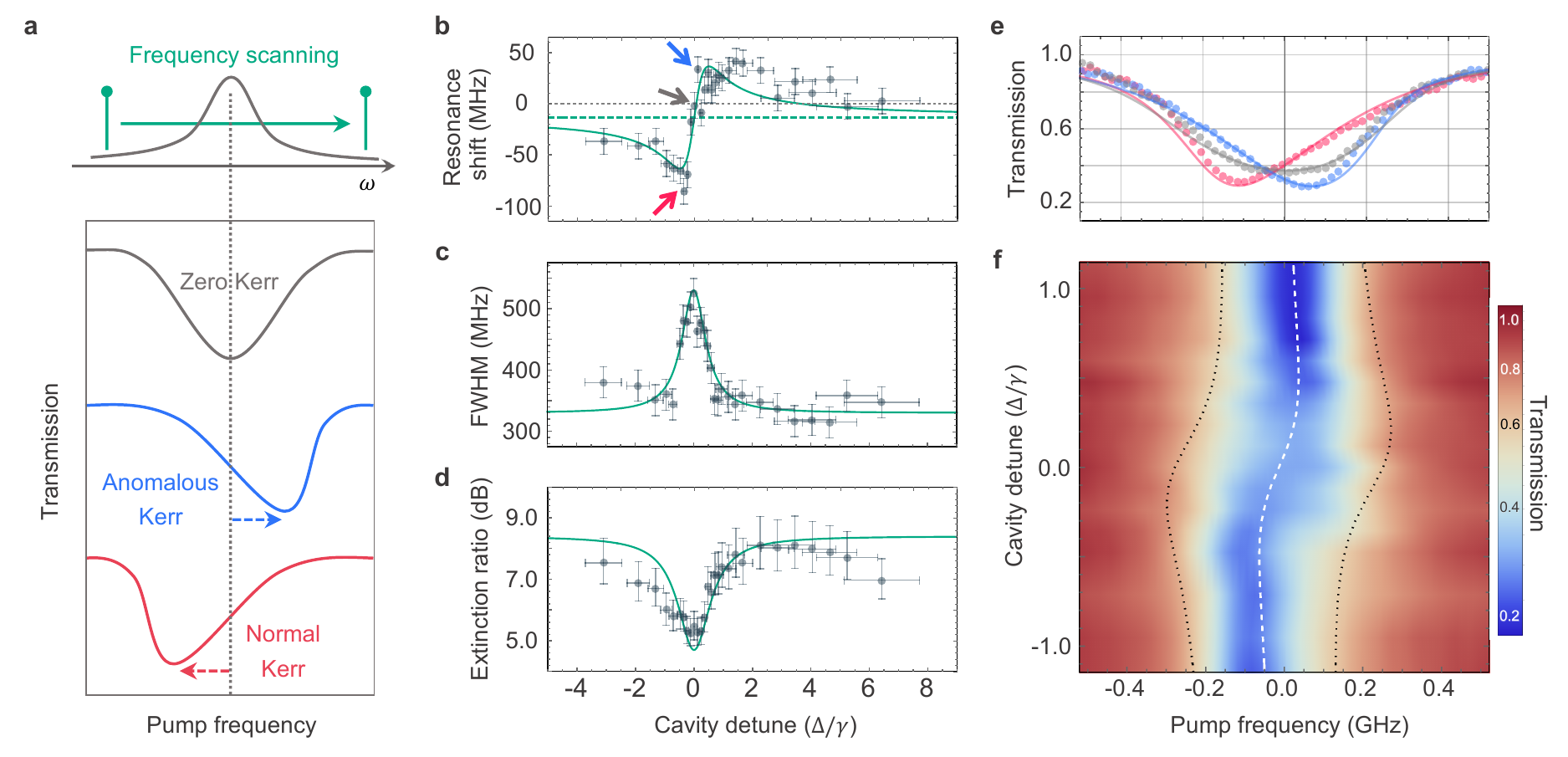}
\caption{\textbf{Anomalous Kerr nonlinearity.} \textbf{a}, Calculated transmission spectrum as the pump frequency is scanned across the resonance. The transmission minimum shifts to higher (blue) and lower (red) frequency with negative and positive Kerr nonlinear coefficients respectively. The transmission spectrum remains symmetric (grey) if the Kerr nonlinear coefficient is zero. 
\textbf{b}, Measured resonance shift dependence on the cavity detune. Gray dashed line: zero frequency shift. Green dashed line: resonance shift induced by the intrinsic Kerr nonlinearity. The blue, red, and grey arrows show the condition to achieve maximum blue shift, maximum red shift, and zero shift, with the corresponding transmission spectrum shown in (\textbf{e}).
\textbf{c}, Measured FWHM dependence on the cavity detune. 
\textbf{d}, Measured extinction ratio dependence on the cavity detune.
\textbf{e}, Measured transmission spectrum with maximum blue shift (blue), maximum red shift (red), and zero shift (grey).
\textbf{f}, Measured transmission spectrum with different cavity detune values. White dashed line: minimum transmission position; dark dashed lines: half maximum position.
Solid lines in \textbf{b}-\textbf{e} are calculated results. Horizontal and vertical error bars in (\textbf{b})-(\textbf{d}) are estimated based on device temperature stability and transmission variance respectively. 
}
\label{fig:Fig2} 
\end{figure*}

In parallel with the intrinsic Kerr nonlinearity, the pump field also undergoes the cascaded second-order nonlinear process consisted of second-harmonic generation and degenerate parametric down-conversion (bottom path in Fig.~\ref{fig:Fig1}a). The pump field is first doubled to generate the second-harmonic field. Then the second-harmonic field drives the degenerate parametric down-conversion to re-generate the pump field. Therefore, the cascaded second-order nonlinearity for the pump field is equivalent to an effective Kerr nonlinearity~\cite{cui2022situ}. In this cascaded processes, the second-order nonlinear strength $g_2$, the second-harmonic resonance linewidth $\gamma$, and the frequency detune $\Delta$ between the pump and second-harmonic resonances collectively determine the effective Kerr nonlinear coefficient. The instantaneous resonance shift is $\delta\omega\propto-\mathrm{Re}\left[\frac{i|g_2|^2}{-i\Delta+\gamma/2}\right] \cdot N_\mathrm{p}$ (Supplementary Information I). With a positive detune $\Delta>0$, the cascaded second-order nonlinearity can lead to the resonance shift towards higher frequency. This is equivalent to self-phase modulation with negative Kerr nonlinear coefficients. The overall resonance shift, after the coherent combination of the intrinsic Kerr and cascaded second-order nonlinear processes, becomes 
\begin{align}
\delta\omega&\propto-\mathrm{Re}\left[g_3+\frac{i|g_2|^2}{-i\Delta+\gamma/2}\right] \cdot N_{\rm p} 
\stackrel{\text{def}}{=} -\mathrm{Re}\left[g_{\rm eff}\right] \cdot N_{\rm p}
\end{align}
with $g_{\rm eff}$ the effective Kerr nonlinear coefficient (Fig.~\ref{fig:Fig1}c). With strong second-order nonlinearity $\xi=|g_2|^2/(g_3\gamma)>1$, the overall resonance shift can be reconfigured towards either lower or higher frequency depending on the cavity detune $\Delta$ (Fig.~\ref{fig:Fig1}b). A completely passive cavity without nonlinear resonance shift can also be obtained at cavity detune $\Delta/\gamma=\frac{1}{2}(\xi\pm\sqrt{\xi^2-1})$. Besides the reconfigurable resonance shift, the cascaded second-order nonlinearity also introduces extra loss for the pump field as power is converted into the second-harmonic field. This is manifested as the imaginary part of the effective Kerr nonlinearity $g_{\rm eff}$.

The reconfigurable Kerr nonlinearity is demonstrated with an aluminum nitride ring cavity \cite{xiong2012aluminum}. Phase-matching condition is realized by geometric dispersion engineering~\cite{zhang2021chip}. Due to self-phase modulation, the transmission spectrum of the pump field deviates from the symmetric Lorentzian shape, and can be used to quantify the Kerr nonlinear coefficient (Fig.~\ref{fig:Fig2}a). The speed to scan the pump frequency is sufficiently fast to avoid the frequency shift induced by the thermal effect (Supplementary Information III). The pump power is maintained below the four-wave-mixing threshold to avoid the generation of optical sidebands. In this low power regime, the frequency difference between the minimum transmission point and the Lorentzian middle point is proportional to the real part of the effective Kerr nonlinear coefficient. The imaginary part can also be obtained through the change in the resonance full-width at half-maximum (FWHM). 

\begin{figure*}[htbp]
\centering
\includegraphics[width=\linewidth]{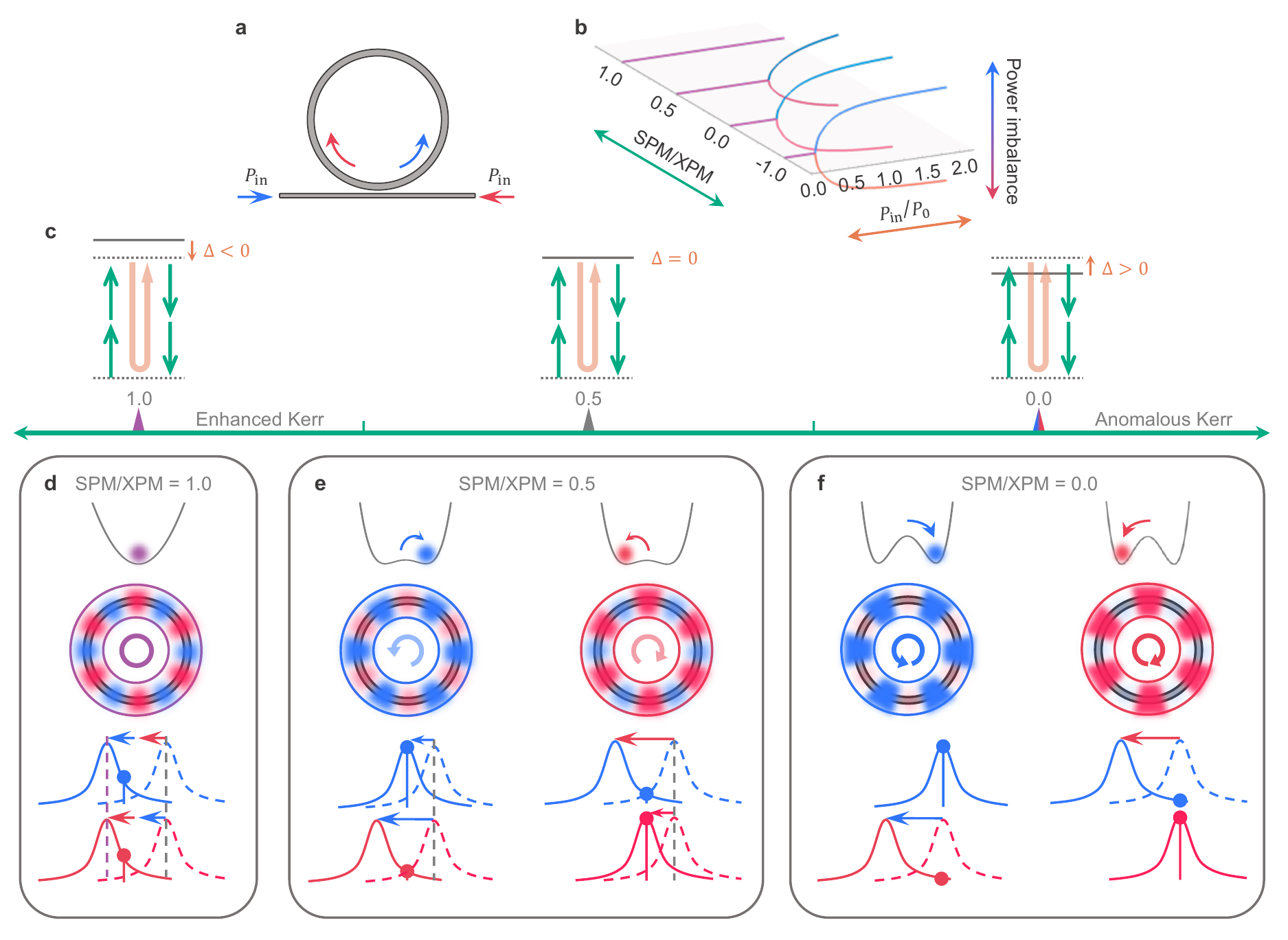}
\caption{\textbf{In-situ control of spontaneous symmetry breaking of photonic chirality}. 
\textbf{a}, Photonic ring cavity with bi-directional pump.
\textbf{b}, Calculated spontaneous symmetry breaking condition with different relative strengths between self-phase modulation (SPM) and cross-phase modulation (XPM). 
\textbf{c}, Cavity detune configuration of the cascaded second-order nonlinearity to realize different SPM/XPM relative strengths.
\textbf{d}, Cavity status under strong pump with SPM/XPM=1. Symmetric state with equal powers is always stable, as the resonance shifts are identical in two directions. 
\textbf{e}, Cavity status under strong pump with SPM/XPM=1/2. The optical field in one direction dominates, as cross-phase modulation introduces twice resonance shift than self-phase modulation.
\textbf{f}, Cavity status under strong pump with SPM/XPM=0. The power difference between different directions become larger as the mismatch between self- and cross-phase modulation increases.}
\label{fig:Fig3} 
\end{figure*}

The cavity detune $\Delta$ is controlled by changing the device temperature, as the pump and second-harmonic resonances have different thermal-optic coefficients (Supplementary Information IV). We can clearly observe the change in the pump transmission spectrum with different cavity detune values (Fig.~\ref{fig:Fig2}f). The resonance shift follows a Fano shape with respect to the cavity detune (Fig.~\ref{fig:Fig2}b). This agrees with the interpretation that the intrinsic Kerr and cascaded second-order nonlinear processes are coherently combined to give rise to the effective Kerr nonlinearity. The pump resonance shows maximum shift towards higher frequency with the cavity detune $\Delta\approx0.5\gamma$, proving the negative self-phase modulation (blue in Fig.~\ref{fig:Fig2}e). By changing the device temperature, we can further reconfigure the pump resonance to shift towards lower frequency and even exhibit no frequency shift (red and grey in Fig.~\ref{fig:Fig2}e). In addition, we also notice that FWHM of the pump resonance follows the Lorentzian shape with respect to the cavity detune (Fig.~\ref{fig:Fig2}c). The largest FWHM is obtained with zero detune $\Delta=0$, in which case maximum power is converted from the pump field to the second-harmonic field. The power conversion can be treated as extra intrinsic loss for the pump resonance. As the cavity is weakly coupled to the bus waveguide with fixed coupling strength, the increase of the intrinsic loss leads to the drop of the resonance extinction ratio (Fig.~\ref{fig:Fig2}d).

\begin{figure*}[htbp]
\centering
\includegraphics[width=\linewidth]{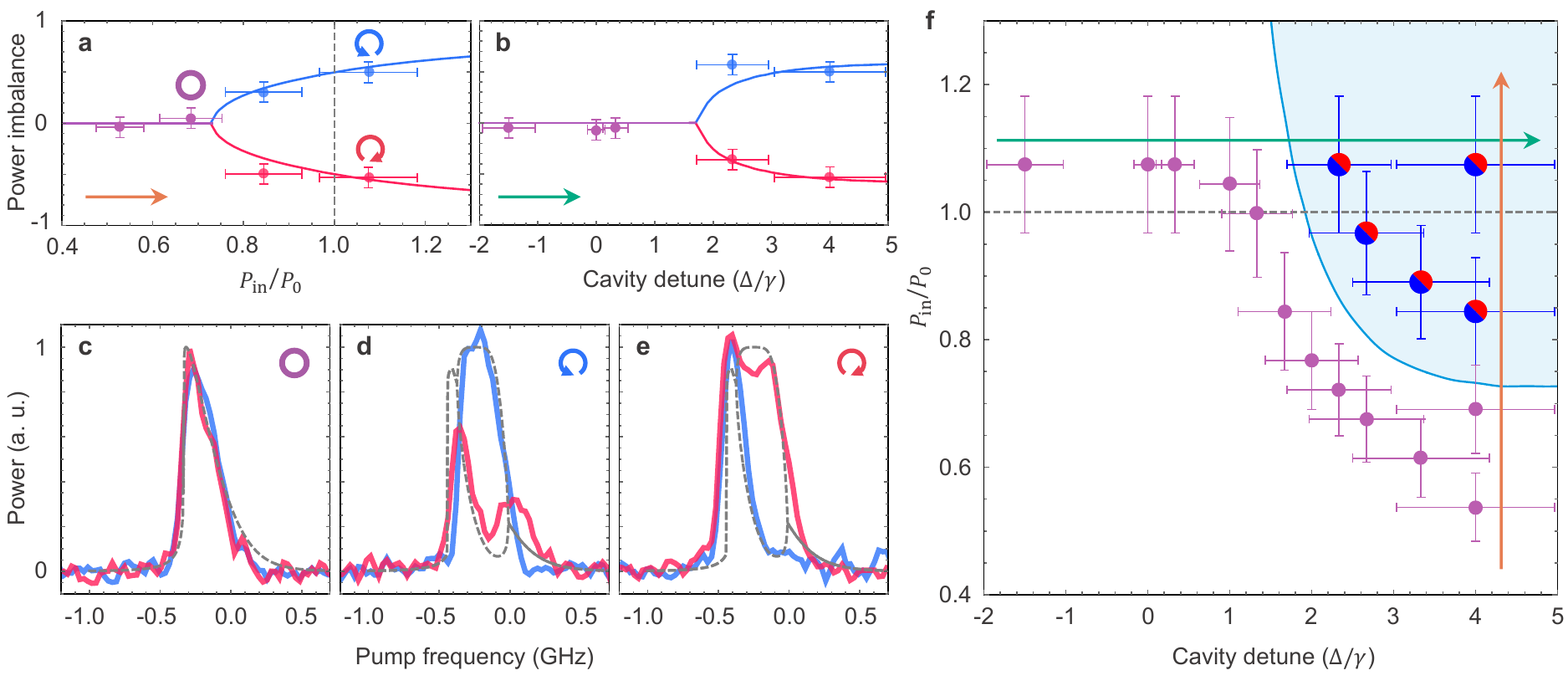}
\caption{\textbf{Control of spontaneous symmetry breaking condition}. 
\textbf{a}, Relative power imbalance with fixed cavity detune $\Delta\approx4\gamma$ and varying pump power $P_{\rm in}$. The threshold power for intrinsic Kerr nonlinearity is marked as dashed black line.
\textbf{b}, Relative power imbalance with varying cavity detune $\Delta$ and fixed pump power $P_{\rm in}\approx1.07P_0$. $P_0\approx51$ mW is the threshold power for spontaneous symmetry breaking with intrinsic Kerr nonlinearity.
\textbf{c-e}, Measured output second-harmonic power in clockwise (blue) and counterclockwise (red) directions with cavity detune $\Delta\approx4\gamma$ and pump power $P_{\rm in}\approx0.7P_0$, $P_{\rm in}\approx1.07P_0$, $P_{\rm in}\approx1.07P_0$, corresponding to the purple, blue, and red circle points in (\textbf{a}) respectively.
\textbf{f}, Cavity status with different cavity detune $\Delta$ and pump power $P_{\rm in}$. Blue solid line: calculated threshold condition for spontaneous symmetry breaking. Shaded blue area: calculated regime with asymmetric power in two directions. Black dashed line: threshold power for intrinsic Kerr nonlienarity. Purple points: measured cavity output with equal power in two directions. Blue/red points: measured cavity output with optical field in one direction dominant. Orange arrow: the same condition change as (\textbf{a}). Green arrow: the same condition change as (\textbf{b}). Power and cavity detune error bars are estimated based on device temperature stability and transmission variance respectively. Additional data for (\textbf{f}) are shown in Supplementary Information V.
}
\label{fig:Fig4} 
\end{figure*}

The key to realize reconfigurable Kerr nonlinearity is the cascaded second-order nonlinear process, which requires the phase matching condition. It provides the unique capability that we can selectively modify the target Kerr nonlinear process without affecting the others. This is achieved by designing the phase matching condition only for the optical fields involved in the target Kerr nonlinear process. For example, we consider the same photonic ring cavity with second-order nonlinearity, but pumped in both clockwise and counter-clockwise directions (Fig.~\ref{fig:Fig3}a). As phase matching condition is only valid for optical fields along one direction, only the self-phase modulation coefficient is modified by the cascaded second-order nonlinearity. To modify the cross-phase modulation, we need sum-frequency generation between counter-propagating fields. This requires zero wave-vector for the second-harmonic field, which cannot be realized. Therefore, cross-phase modulation coefficient remains unchanged. 

The bi-directional configuration is of critical importance for numerous photonic applications including optical gyroscope~\cite{lai2019observation}, single-particle detection~\cite{chen2017exceptional}, and soliton referencing~\cite{yang2017counter}. It is recently realized that spontaneous symmetry breaking of photonic chirality can also be introduced in this bi-directional configuration due to the imbalance between self- and cross-phase modulation~\cite{del2017symmetry,cao2017experimental,garbin2020asymmetric,cao2020reconfigurable,silver2021nonlinear}. For intrinsic Kerr nonlinearity, the ratio between self- and cross-phase modulation has a fixed value of $1/2$ in solid-state materials~\cite{boyd2020nonlinear}. The optical field along one direction causes twice resonance shift for the opposite direction. Therefore, the optical field with smaller power will experience larger resonance shift, which further suppresses its amplitude due to the larger mismatch between the input frequency and cavity resonance. With this positive feedback mechanism, the symmetric state with equal power in both directions becomes unstable even the input pumps have the same power (Supplementary Information II). Then the optical field in one direction becomes stronger, showing spontaneous chiral behavior (Fig.~\ref{fig:Fig3}e).

Obviously, the condition for spontaneous symmetry breaking is directly determined by the relative strength between self- and cross-phase modulation. With the capability to selectively control self-phase modulation without affecting cross-phase modulation, we can engineer a single cavity to exhibit different behaviors for spontaneous symmetry breaking (Fig.~\ref{fig:Fig3}b-f). If we enhance the self-phase modulation to the same strength of cross-phase modulation, spontaneous symmetry breaking can be prohibited as optical fields in both directions will experience the same resonance shift regardless of the power distribution (Fig.~\ref{fig:Fig3}d). If we suppress the self-phase modulation, spontaneous symmetry breaking can be realized more easily as the resonance shift difference is larger with lower pump power (Fig.~\ref{fig:Fig3}f).

To observe spontaneous symmetry breaking, optical fields with identical frequency and power are coupled into the cavity in both directions (Supplementary Information III). When the optical frequency is scanned across the pump resonance, we monitor the output second-harmonic power, which is proportional to the square of the intra-cavity pump photon number. The photonic chirality $C$ is defined as the relative intra-cavity power imbalance
\begin{align}
C=\frac{(N_\mathrm{p}^\mathrm{CCW}-N_\mathrm{p}^\mathrm{CW})}{(N_\mathrm{p}^\mathrm{CCW}+N_\mathrm{p}^\mathrm{CW})}
\end{align}
with $N_\mathrm{p}^\mathrm{CCW}$ and $N_\mathrm{p}^\mathrm{CW}$ the intra-cavity pump photon number along clockwise and counter-clockwise directions respectively. Larger power imbalance indicates stronger spontaneous symmetry breaking of photonic chirality. 

We first vary the pump power while fixing the cavity detune at which the self-phase modulation is suppressed (Fig.~\ref{fig:Fig4}a). With low pump power, the optical fields have the same amplitude in both directions (Fig.~\ref{fig:Fig4}c). By increasing the pump power, we observe the spontaneous symmetry breaking with the optical field in one direction significantly stronger than the other (Fig.~\ref{fig:Fig4}d\&e). Due to the suppression of the self-phase modulation, the threshold power for spontaneous symmetry breaking is significantly lower compared to the intrinsic Kerr nonlinearity case (Fig.~\ref{fig:Fig4}a). 
Next, we fix the pump power, and vary the cavity detune $\Delta$ for the cascaded second-order nonlinearity (Fig.~\ref{fig:Fig4}b). With negative cavity detune, the self-phase modulation is enhanced. Therefore, the optical fields along both directions remain the same. By increasing the cavity detune, the imbalance between self- and cross-phase modulation becomes larger, leading to the spontaneous symmetry breaking. 
As shown in Fig.~\ref{fig:Fig4}f, the spontaneous symmetry breaking condition is collectively determined by the pump power and cavity detune. The threshold pump power can be continuously tuned by the cavity detune. Therefore, we can either prevent or facilitate the occurrence of spontaneous symmetry breaking by increasing or decreasing the threshold pump power.

In summary, we have demonstrated the capability to control the spontaneous symmetry breaking of photonic chirality. This is achieved by directly modifying Kerr nonlinear interactions with the cascaded second-order nonlinearity. Beyond the fixed intrinsic Kerr nonlinearity, we can realize reconfigurable anomalous Kerr effects, including negative self-phase modulation and strength tuning between competing Kerr nonlinear processes. This flexible mechanism can find broad applications in photonic technologies based on Kerr nonlinearity, such as frequency comb generation, quantum state generation, and frequency conversion. Moreover, the control of spontaneous symmetry breaking condition can provide novel insights to the study of nonlinear dynamics in broad fields ranging from condensed-matter physics to nanomechanics.

\subsection{Acknowledgments}
CC, LZ, and LF acknowledge the support from U.S. Department of Energy, Office of Advanced Scientific Computing Research (Field Work Proposal ERKJ355); Office of Naval Research (N00014-19-1-2190); National Science Foundation (ECCS-1842559). Device fabrication is performed in the OSC cleanrooms at the University of Arizona and the cleanroom of Arizona State University. 

\subsection{Author Contributions Statement}
CC and LF conceived the idea, designed the experiment, and wrote the manuscript. LZ fabricated the device. CC derived the theoretical model and operated the experiment. LF supervised the work.

\subsection{Competing Interests Statement}
The authors declare no competing interests.

\bibliography{Ref}

\bibliographystyle{apsrev4-2}

\subsection{Methods}

\textbf{Device fabrication and parameters.} Devices were fabricated from 1-$\mu$m aluminium nitride grown on sapphire substrates using metalorganic vapour-phase epitaxy (MOCVD). FOx-16 resist was used for patterning photonic circuits with electron-beam lithography. After development with tetramethylammonium hydroxide (TMAH), plasma etching with Cl2/BCl3/Ar was used to transfer the pattern to the aluminium nitride layer. Finally, SiO2 cladding was deposited by plasma-enhanced chemical vapour deposition (PECVD). 

The ring cavity has a radius of 60~$\mu$m and width of 1.1~$\mu$m. The fundamental and second-harmonic resonances have quality factors of $5.82\times10^5$ near 1560~nm and $1.83\times10^5$ near 780~nm respectively. The intrinsic Kerr nonlinear coefficient is estimated as $g_3\approx2\pi\times2.0$~Hz. This gives the threshold power of spontaneous symmetry breaking with intrinsic Kerr nonlinearity $P_{\rm 0}\approx 51$~mW. 
The second-order nonlinearity strength is estimated as $g_2\approx2\pi\times122$~kHz. This leads to relative strength between cascaded and intrinsic Kerr nonlinearity $\xi=|g_2|^2/(g_3\gamma)\approx 3.7$ (Supplementary Information IV).

\textbf{Measurement setup.} Self-phase modulation is measured with 10~mW on-chip pump power, which is strong enough to introduce significant frequency shift without generating optical sidebands by four-wave mixing. Fast frequency scan is realized with single-sideband electro-optic modulation. The time to scan across the resonance is less than 20 ns (scanning rate 36~GHz/$\mu$s). The transmitted light is detected with a 1~GHz IR photodetector. Therefore, thermal-optic effect (millisecond timescale) is excluded in the measurement. 

The spontaneous symmetry breaking is observed by measuring the output second-harmonic field under bi-directional pump. We use a lower scanning rate (scanning rate 6~GHz/$\mu$s), as the visible photodetector has a smaller bandwidth (125~MHz).
The complete measurement setup is shown in Supplementary Information III.

\subsection{Data availability}
All the data supporting the plots within this paper and the code are available from the corresponding author upon reasonable request.

\end{document}